\documentclass[dvips]{elsart}
\usepackage{epsfig}
\usepackage{times}
\usepackage{graphicx}
\usepackage{caption2}
\usepackage{placeins}
\usepackage{rotate}
\begin{document}
\begin{frontmatter}
%
\title{\boldmath Calculations of neutrino spectra with a program
based on the GEANT library.}
\author[INR]               {M.~Kirsanov}

\address[INR] {Inst. Nucl. Research, INR Moscow, Russia, Mikhail.Kirsanov@cern.ch}           

\clearpage
\begin{abstract}

The methods of neutrino spectra calculations with GEANT library and
the peculiarities of such calculations for different neutrino beams
are summarized. The formula expressing $K^0$ yields from a neutrino
target through the $K^+$ and $K^-$ yields from the same target is
proposed.
\end{abstract}
\end{frontmatter}

\newpage

\section{Introduction}
\label{sec:introduction}

\hspace*{0.5cm}
In 1988 - 1989 the new generation of neutrino experiments started
at IHEP (Protvino). It was a result of the physical start of the new
IHEP-JINR neutrino detector, increased intensity of the IHEP 70 - GeV
proton accelerator and somewhat changed situation in the neutrino physics,
which required better accuracy of measurements and inspired interest to
experiments with electron neutrinos. \\
\hspace*{0.5cm}
Most of these experiments were carried out with a nonstandard mode of
work of the neutrino facility. It concerns the beam-dump experiment,
the experiment with focusing system switched off and the new experiment
with short decay base. These experiments required a new program for
neutrino spectra calculations which could take into account additional
effects that become important in these modes of the neutrino beam
formation, for example hadronic interactions in the elements of the
neutrino channel other that the neutrino target. \\
\hspace*{0.5cm}
For this purpose a program based on the GEANT library \cite{geant}
was chosen. It has the following advantages:
\begin{itemize}
\item
developped, multiply checked and supported package for a simulation
of passage of particles through matter, including hadron showers;
\item
the way of describing a geometry is sufficiently developped and 
familiar to most of physicists.
\end{itemize}
\hspace*{0.5cm}
The disadvantage of the full GEANT - like simulations of processes
in the neutrino channel is that it is rather time consuming. This
required usage of tricks that would increase the number of useful
decays, but in spite of them, only after appearance of sufficiently
powerful computers an accurate calculation of neutrino spectra by such
a program became possible.

\section{The program for the neutrino spectra calculation}
\label{sec:setup}

\hspace*{0.5cm}
In the GEANT library the procedures for random decays of particles
in flight are provided. If GEANT accomplishes a decay, the particle
is stopped (no more traced). Thus, a particle can decay only once,
which required the simulation of many particles and a lot of CPU
time. For this reason routines different from the ones of the GEANT
library were used for the simulation of decays. These routines are
called several times on every step of a particle. The GEANT table
of branching ratios was changed (some branching ratios are zeroed),
so that GEANT does not accomplish decays to neutrinos of the type
for which the spectra are being calculated. To take into account
correctly the zeroed branching ratios the GEANT routine GDECAY is
slightly changed. Namely, if no decay mode is chosen (it is
now possible since the sum of branching ratios is not equal to unity),
a particle is not stopped. Thus, a particle makes more steps, which
also improves the statistical accuracy of the calculations. The effect
of decreasing number of particles along the beam due to a finite
lifetime is taken into account through weights ascribed to particles. \\
\hspace*{0.5cm}
Unfortunately, although all these tricks certainly decrease the overall
statistical errors of calculations, they do not decrease by the same
factor all components of these errors. For example, the fluctuations
due to hadron showers are not decreased. This makes very difficult a
direct estimation of statistical uncertainties of spectra.
The only obvious way to estimate them is to compare results of
several runs of the program. This consideration also suggests that
there is no sense in very small steps or many decays in one step. \\
\hspace*{0.5cm}
Some branching ratios, such as the branching ratios to hadrons, are
not changed in order to simulate more exactly the beam composition
(for example if a kaon decays to pions, these pions are also traced
later). For the same purpose and in order to take into account
decays of muons during the calculation of electron neutrino
spectra the program has 2 modes of work:
\begin{enumerate}
\item
The calculation of muon neutrino and antineutrino spectra. The
branching ratios of decays to electron neutrinos are not changed
since in some of them charged pions are produced.
\item
The calculation of electron neutrino and antineutrino spectra.
The branching ratios of decays to muons are not zeroed.
\end{enumerate}
\hspace*{0.5cm}
In each of these modes it is possible to simulate directly hadron
showers in the neutrino target induced by initial protons or
(not for the beam dump) to use parameterized yields of particles
from the standard neutrino targets. \\
\hspace*{0.5cm}
As most of programs using GEANT library the program consists
of the following parts:
\begin{enumerate}
\item
The initialization. Here beside usual GEANT calls the GEANT table of
branching ratios is changed.
\item
The geometry definition. It is a usual GEANT geometry definition
\cite{geant} except that some volumes should have standard numbers
(the target, lenses and some volumes for the calculation of control
distributions). Besides, the constants for the spectra calculations,
such as a distance to a possible detector and its maximal sizes,
are set.
\item
The routine for the action before each event. Here either a proton
is put into the GEANT initial bank (direct calculation) or several
hadrons with random momentum and angle (parameterization).
\item
The routine for the action after each step. It calls the two
main routines of the program: \\
a) The routine that ascribes a weight to a new particle or calculates
a weight after each step. A weight 1 is ascribed to primary protons.
In the mode of work with a parameterization a weight to primary
hadrons at their first appearance is ascribed according to this
parameterization. For other particles, at their first appearance,
their creation vertex is searched for in the array of vertices
(this is not in the standard GEANT banks and common blocks, see
below). From this array the initial weight, the type and momentum
of a parent particle, the spin direction (for a muon) are
taken. If there is a correction factor for the process of creation
of this particle by its parent particle the weight is corrected.
After each step the weight is decreased according to the life
time of particle divided by the sum of all zeroed branching ratios. \\
b) The calculation of the neutrino spectra. For two body decays
the limits on the angles in the center of mass system are calculated,
so that neutrinos from a decay can be inside the maximal detector
volume. For three body decays, if a neutrino energy is above
threshold and its direction does not hit the detector maximal
volume, its azimuthal angle is resimulated withing the limits that
contain the maximal detector volume. The probability of a decay
in case of additional limits is multiplied by the corresponding phase
space factor. The matrix element of decays is taken into account. For
decays of muons it depends on the spin direction. \\
\hspace*{0.5cm}
The probabilities of decays are put into double precision arrays
that correspond and will be rewritten in the end of run into
one- and two-dimensional histograms (energy - radius, energy - angle).
\item
The routine for the action after each act of creation of new particles.
As usual, particles below cut and those that do not contribute to
the neutrino spectra, are ignored. Here the array of vertices mentioned
above is filled
\end{enumerate}

\section{The spectra of hadrons.}

\begin{figure}[h]
\begin{center}
   \mbox{
     \epsfig{file=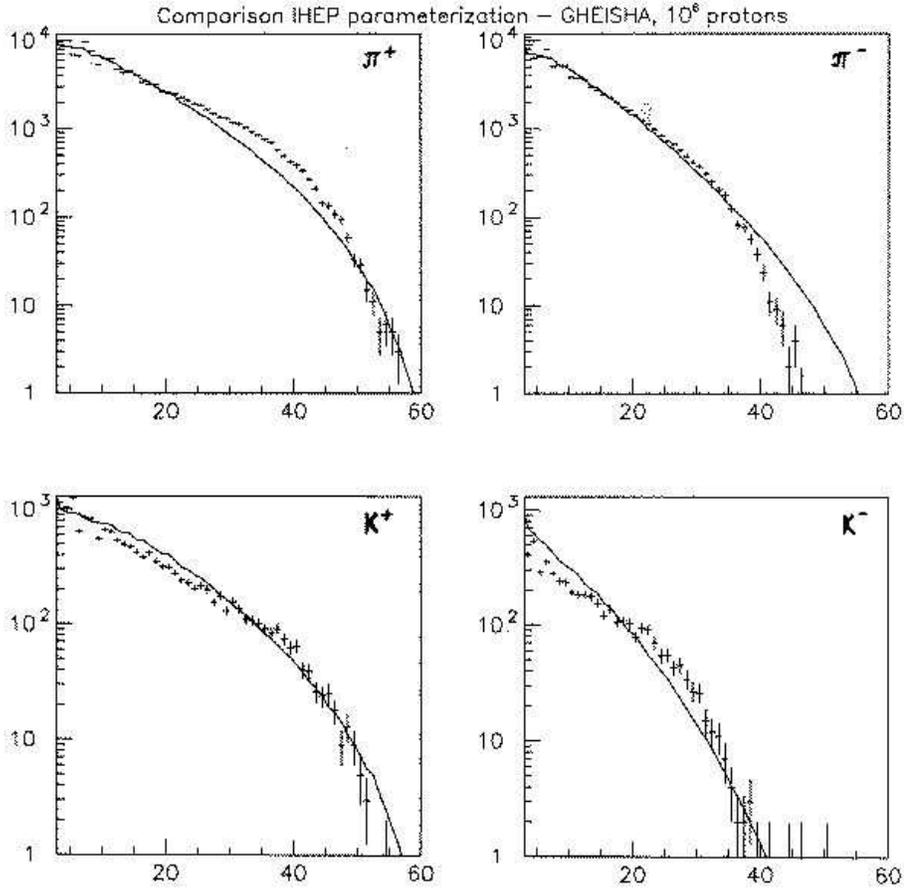,width=130mm}}
     \caption{The spectra of secondary particles from the thick
   aluminium target exposed to the beam of 67 - GeV protons, integral
   over the angle range 0 - 30 mrad. GEANT 321 with GHEISHA
   (CERN library 95a) - points with errors, line - measurements \cite{yields}}
      \label{fig:gheisha}
   \end{center}
\end{figure}

\begin{figure}[h]
\begin{center}
   \mbox{
     \epsfig{file=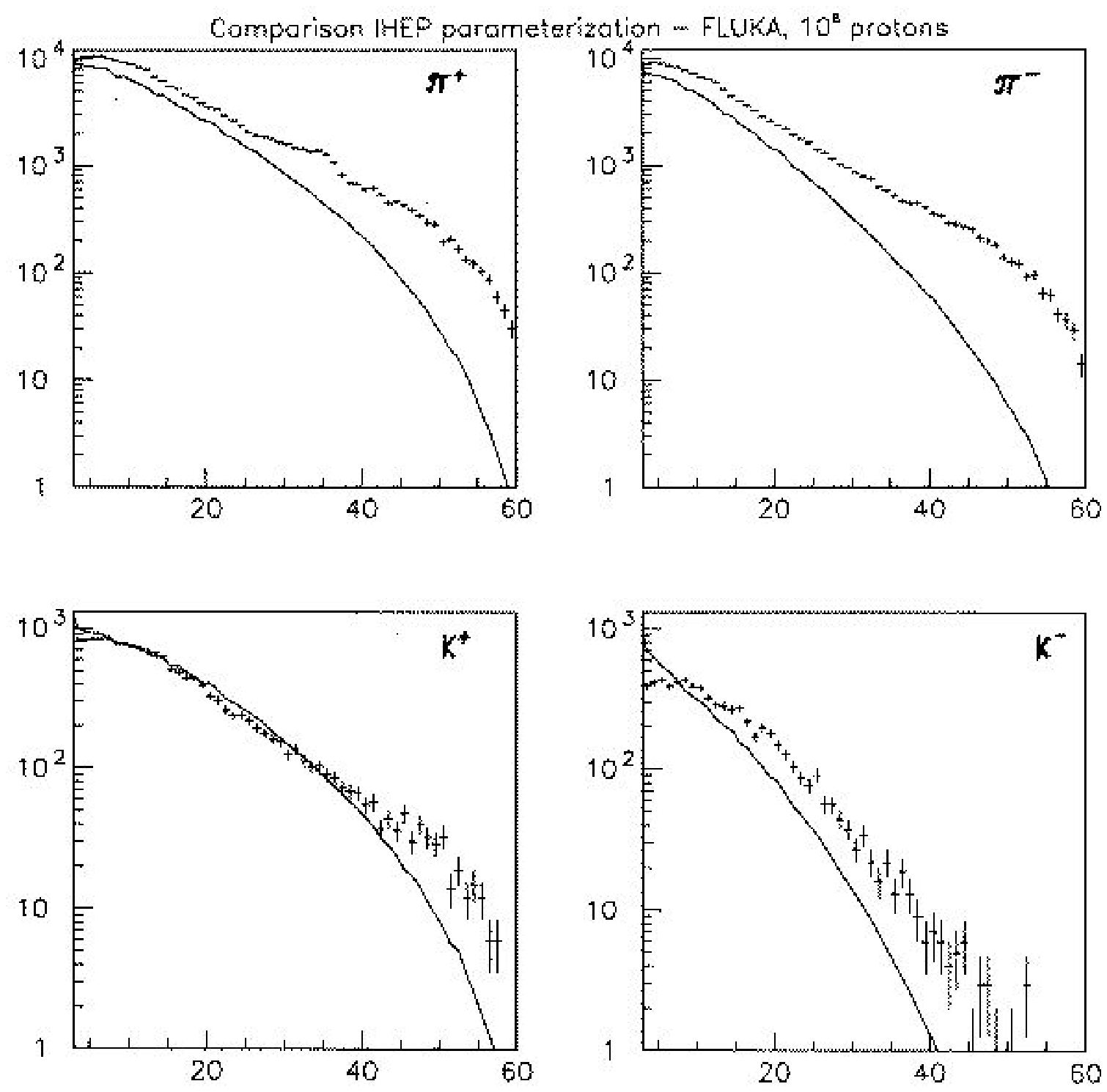,width=130mm}}
     \caption{The spectra of secondary particles from the thick
   aluminium target exposed to the beam of 67 - GeV protons, integral
   over the angle range 0 - 30 mrad. GEANT 321 with GEANT - FLUKA
   (CERN library 95a) - points with errors, line - measurements \cite{yields}}
      \label{fig:fluka}
   \end{center}
\end{figure}

\hspace*{0.5cm}
The first calculations of WBB spectra were made in the mode with
the full simulation of the hadron shower in the neutrino target,
which is a 60 cm long aluminium cylinder. Although the final
calculations were made using the parameterization, it is interesting
to compare the GEANT yields with the ones measured in \cite{yields}.
These measurements are rather detailed, they include secondary
particles from the primary beam of 67 - GeV protons in the momentum
range from 5 - 7 GeV to 60 GeV and the angle range from 0 to 50 mrad,
subdivided into energy bins of 1 GeV and angle bins of 2 mrad. \\
\hspace*{0.5cm}
The comparison for GEANT 321 with GHEISHA is shown in Fig.~\ref{fig:gheisha}. \\
\hspace*{0.5cm}
The comparison for GEANT 321 with FLUKA is shown in Fig.~\ref{fig:fluka}.

\begin{figure}[h]
\begin{center}
   \mbox{
     \epsfig{file=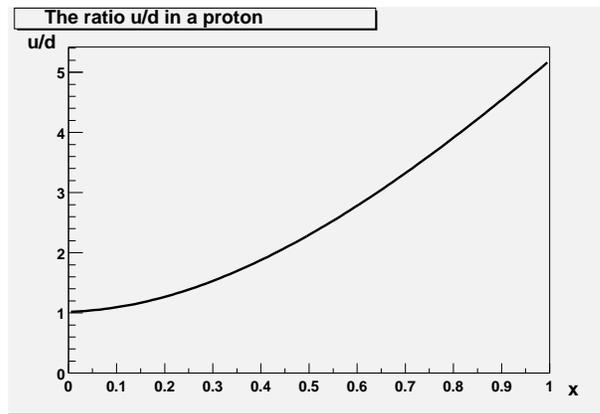,width=80mm}}
     \caption{The ratio of the distributions u/d in a proton \cite{utod}}
      \label{fig:utod}
   \end{center}
\end{figure}

\hspace*{0.5cm}
From these fugures one can see that GEANT does not simulate correctly
the yields of hadrons, the difference reaching a factor of 2 for the
case of GHEISHA. For this reason in the mode of work with a full
simulation the correction by weights (previous section, 4a) should
be used. \\
\hspace*{0.5cm}
Special attention should be paid to the spectrum of $K^0$, for which
there are no reliable measurements at 70 GeV. Often it was assumed
that there is the following relation between the charged and neutral
K mesons:
\begin{equation}
K^0_L = K^0_S = \frac{K^++K^-}{2}
\end{equation}
\hspace*{0.5cm}
It is clear from the following considerations that the relation
should be different. We are usually interested in spectra of rather
fast particles, which means that they should be "leading" particles,
i. e. should contain a quark from the incident proton. Simple
counting of quarks shows that the spectrum of $K^0$ should be closer
to the spectrum of $K^-$ than to the spectrum of $K^+$. \\
\hspace*{0.5cm}
In order to find the best description the following measurements were
analysed: the $K^0$ measurement at FNAL \cite{k0fnal} at 300 GeV,
the charged kaon measurements at CERN at 400 GeV \cite{kcern} and the
charged kaon measurements \cite{yields} at IHEP (Protvino) at 67 GeV. All
measurements were made on thin targets. The method \cite{method} of
comparison of spectra at different energies was used. It uses the
variable $x_R = E^*/E^*_{max}$, the ratio of the energy in the
center of mass system to the maximal energy at a given $p_t$. The best
description was obtained with the following formula:
\begin{equation}
K^0_L = K^0_S = \frac{K^++(2n-1)K^-}{2n}
\end{equation}
where $n$ is the ratio of the distributions u/d in a proton. This
value is a function of a scaling variable $x$, which we substitute
here by $x_R$, which does not introduce a large error since the
function does not change rapidly with $x$. It can be found for example
in \cite{utod}. $n(x_R)$ changes from 1 to 5 as $x_R$ changes from 0 to 1
(Fig.~\ref{fig:utod}). The following approximation was used afterwards
in calculations of spectra:
\begin{equation}
n = -1.435x^3_R + 5.3883x^2_R + 0.22269x_R + 1.0189
\end{equation}

\begin{figure}[h]
\begin{center}
   \mbox{
     \epsfig{file=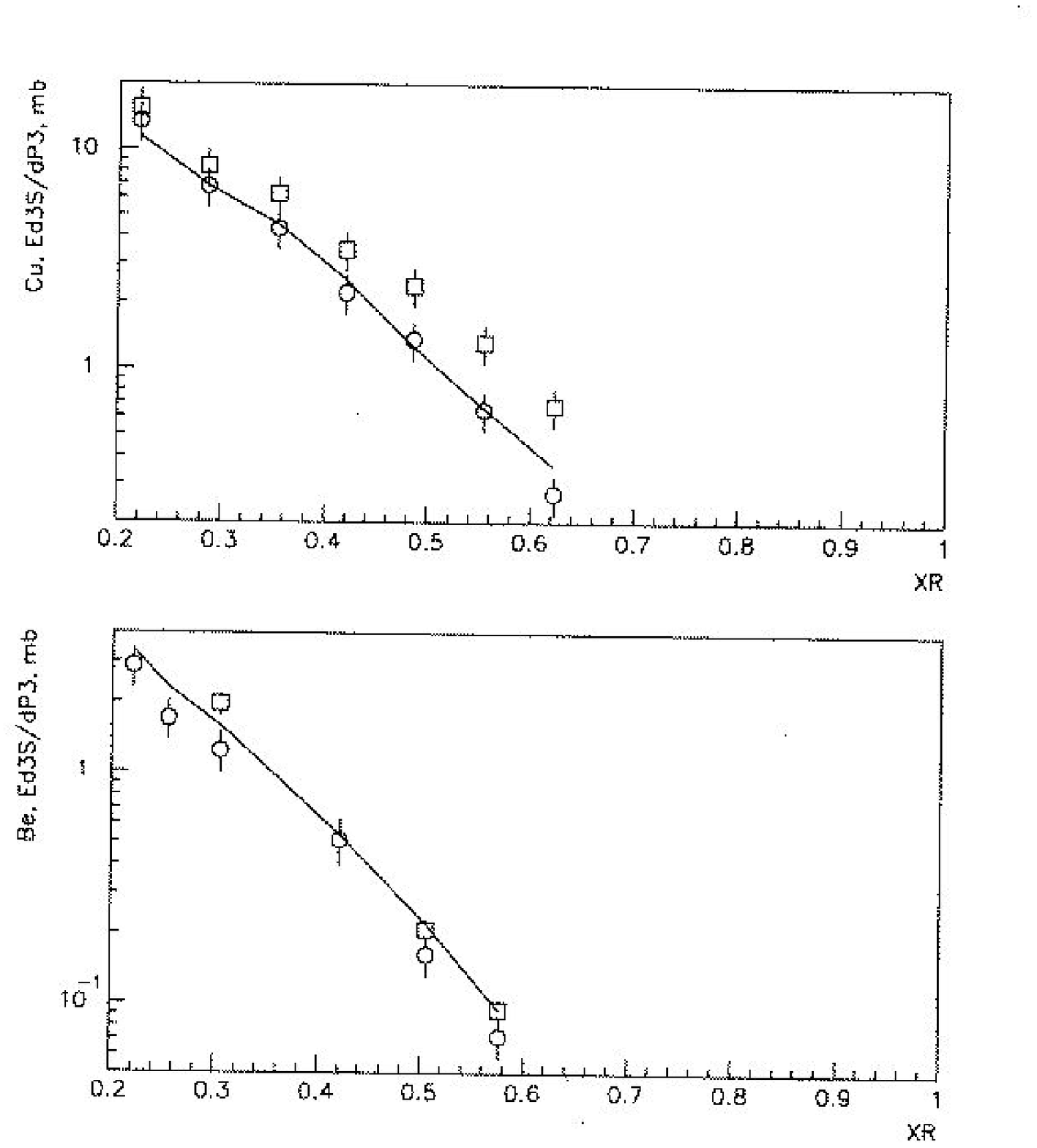,width=120mm}}
     \caption{{\bf a)} The line - $K^0$ spectrum measured at FNAL; circles
    - $K^0$ spectrum calculated according to the formula (2) from the charged
    kaon spectra measured at IHEP on a copper target; squares - $K^0$
    spectrum calculated according to the traditional formula (1).
    {\bf b)} The line - $K^0$ spectrum measured at FNAL; circles  - $K^0$
    spectrum calculated according to the formula (2) from the charged
    kaon spectra measured at IHEP on a berillium target; squares - $K^0$
    spectrum calculated according to the formula (2) from the charged
    kaon spectra measured at CERN on a berillium target. All spectra
    are at a fixed $p_t$ relevant for neutrino spectra. Uncertainties
    of neutral kaon spectra are shown, uncertaintied of charged kaon
    spectra are not shown.}
      \label{fig:k0}
   \end{center}
\end{figure}

\hspace*{0.5cm}
The uncertainty of formula (2) is estimated to be 15\%, which is a
quadratic sum of the uncertainty of the FNAL measurements (10\%),
the uncertainty of the scaling from 70 to 400 GeV (10\%) and
a smaller uncertainty of the CERN measurements. \\
\hspace*{0.5cm}
In Fig.~\ref{fig:k0} one can see that the formula (2) is in a much
better agreement with the copper data than $(K^++K^-)/2$. The agreement
of the formula (2) on berillium is also quite reasonable. Apparently
the measurements \cite{yields} on berillium slightly underestimate
the yields.

\section{The beam dump neutrino spectra.}

\begin{figure}[h]
\begin{center}
   \mbox{
     \epsfig{file=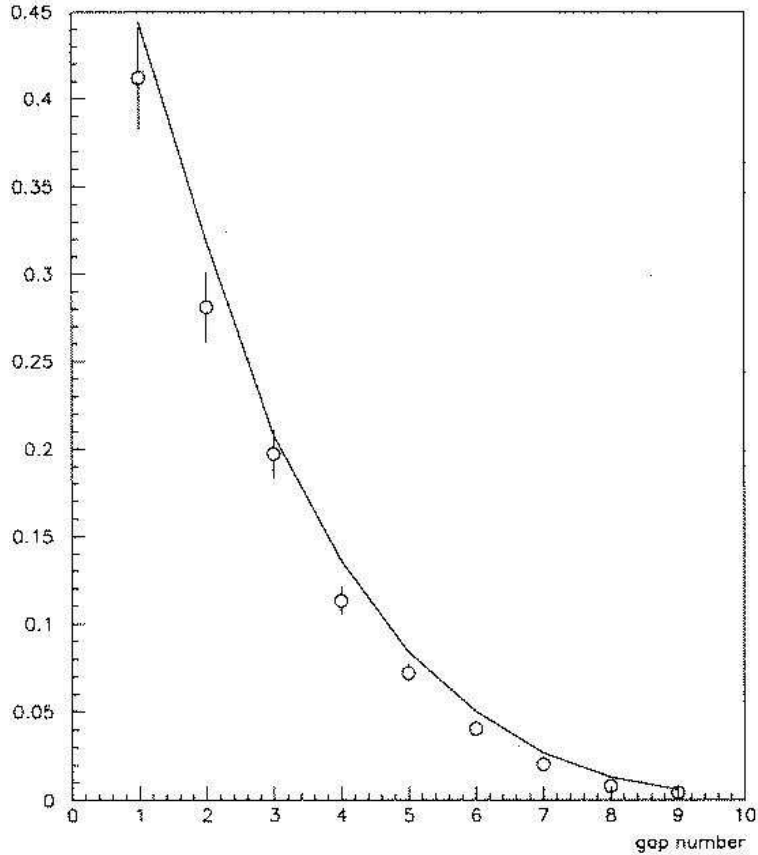,width=100mm}}
     \caption{Beam-dump muon fluxes $\times 1000$, integrals inside the
              circle with a radius 70 cm, per one incident proton. Line is the
              result of the calculation.}
      \label{fig:bdmu}
   \end{center}
\end{figure}

\hspace*{0.5cm}
Practically the only possible way of spectra calculation for
beam dump experiments is the mode with a full simulation of
hadron showers in the target. It is difficult here to use
parameterizations. However, at least the GEANT (GHEISHA) yields of particles
from the first interaction of protons were compared with the results
of measurements on the thin copper target \cite{yields}. The difference
was similar to that seen in Fig.~\ref{fig:gheisha}. A corresponding
correction was used in the calculations of spectra (see section 2, 4a). \\
\hspace*{0.5cm}
The systematic uncertainties of the calculated spectra were determined
by the systematic uncertainties of the measurements \cite{yields}.
Possible uncertaintied in the simulation of the secondary interactions
by GEANT were suppressed since corresponding secondary particles
contribute 20 - 30\% to the neutrino spectra interesting
for the beam dump experiment. These contributions were so small due to the
fact that the experiment \cite{bd} was devoted to the search of hard
neutrinos from the charm production which required rather high energy cuts. \\
\hspace*{0.5cm}
It is possible to reduce these uncertainties for the electron neutrino
and antineutrino spectra. For this purpose the reconstructed muon neutrino
events were used. The high energy region of the muon neutrino and antineutrino
spectra is dominated by decays of charged kaons (the contribution
from charm estimated from the beam-dump extrapolation method was small).
For this reason, comparing the number of reconstructed muon
neutrino events with the calculated one it is possible to check the
charged kaon spectra and to reduce systematic uncertainties. \\
\hspace*{0.5cm}
The final accuracy of the spectra calculations were 10\% for the
electron neutrinos and 16\% for the electron antineutrinos. \\
\hspace*{0.5cm}
Besides, for the beam-dump experiment such important value as a
ratio of effective densities of the two targets can be calculated
with a very good accuracy. Since the dimensions of the half-density
target were finite and there were constructive gaps, it is not
exactly 2 as it was designed. Furthermore, this ratio for muon
neutrinos is different from that for electron neutrinos since for
the latter there is a contribution from $K^0_S$. For such
calculations mainly a correct simulation of the shower shape is
needed, which is reproduced by GEANT more reliably (this was
checked, for example, during the calibration of the neutrino
detector). The calculated ratio of the effective densities
is 2.0 for muon neutrinos and 1.9 for electron neutrinos
(accuracy 1\%, mainly statistical). \\
\hspace*{0.5cm}
For additional control of the beam and spectra during the beam-dump
experiment the muon fluxes in 9 gaps of the muon filter were
measured \cite{mufl}. In Fig.~\ref{fig:bdmu} the calculated fluxes are
compared with the measured ones (the accuracy of measurements 3\%, to which
the uncertainty of the prompt muon pair production subtraction 7\%
should be added). Unfortunately, such control does not allow to improve
the accuracy of the electron neutrino spectra calculations since
the main source of muons (more than 80\%) is the decay of pions.

\section{The neutrino beam with the focusing system turned off.}

\begin{figure}[h]
\begin{center}
   \mbox{
     \epsfig{file=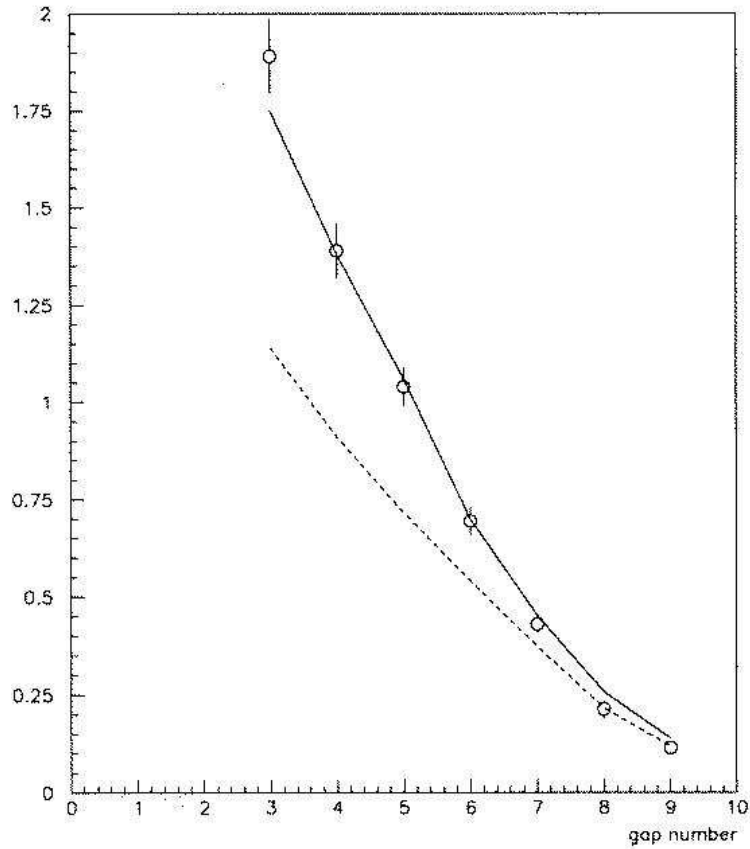,width=100mm}}
     \caption{Muon fluxes $\times 1000$ for the neutrino beam with focusing
              system off. Solid line - calculations, circles - measurements.
              dashed line - calculations without secondary effects.}
      \label{fig:nofocmu}
   \end{center}
\end{figure}

\hspace*{0.5cm}
Two rather long runs of the IHEP-JINR neutrino detector were
carried out with the neutrino channel focusing system turned off.
Such mode of work has the advantage of a higher fraction of
electron neutrinos, sufficient to perform some studies. These
runs were also used for the measurement of the total cross
section of neutrino and antineutrino interactions (simultaneously).
The corresponding total neutrino fluxes are significantly lower
than with the focusing system turned on, but the loss of statistics
for the neutrino detector was not dramatic because due to the lack
of a resonant beam extraction (only a fast extraction is
available in IHEP Protvino) the event aquisition rate is limited
by the maximal allowed occupancy of the neutrino detector. In these
measurements it is also important that neutrino and antineutrino
spectra are less steep, which decreases the effects of smearing. \\
\hspace*{0.5cm}
Since in this neutrino beam (as well as in all the beams described
below) the standard neutrino target was used, the calculations were
accomplished in the mode of work of the program that uses the
parameterization of secondary particle spectra from the target.
But the data of \cite{yields} could not be used directly since the
measurements \cite{yields} were carried out with a 6 cm diameter
aluminuim target, whereas in the neutrino runs a 1 cm diameter
target was used. The neutrino spectra in this case are better
(harder), but the parameterization should be corrected. The correction
was obtained in the following way: the program was run in the mode
of full shower simulation and the correction was obtained by dividing
the corresponding yields of hadrons from the targets. \\
\hspace*{0.5cm}
In the mode of work with the parameterization the GEANT correction
procedure mentioned in section 2, 4a, is still needed since part
of protons traverse the target without interaction and interact in the lenses,
which is simulated by GEANT. What is significantly better in this mode
of work is that statistical errors at high energy are much lower since
the unweighted spectrum of hadrons from a target to be traced by GEANT is
flat (is not a function of momentum). \\
\hspace*{0.5cm}
It turned out that without focusing the secondary effects, such
as interactions of hadrons in the channel elements are much more
important than with normal focusing. This happens because a bigger
fraction of particles hit the walls of the decay tunnel. In addition,
when the focusing system is turned on, the secondary particles from
the interactions in the neutrino channel elements are not focused or
are not in optimal conditions for a focusing. For this reason they
contribute less to the neutrino spectra. \\
\hspace*{0.5cm}
Only after calculations with a program based on GEANT the agreement
between the calculated and measured \cite{mufl} muon fluxes was obtained.
This is shown in Fig.~\ref{fig:nofocmu}. \\

\hspace*{0.5cm}
The systematic errors of the muon neutrino spectra in the first
approximation are those of \cite{yields}, but they are additionally
controlled, mainly below 10 GeV (pionic part), by the muon flux
measurements (the accuracy of the measurements about 3\%). \\
\hspace*{0.5cm}
In the electron neutrino spectrum for this neutrino channel
a contribution from muon decays is significant. For the calculation
of this contribution it is important to take into account the
polarization of muons. These decays increase the electron neutrino
spectrum by up to 20\% at 3 GeV (thay give softer neutrinos so this
difference is the smaller the higher is the energy). This effect is
even more important for the neutrino beam with focusing. \\

\section{The neutrino beam with the focusing system on.}

\begin{figure}[h]

\begin{center}
   \mbox{
     \epsfig{file=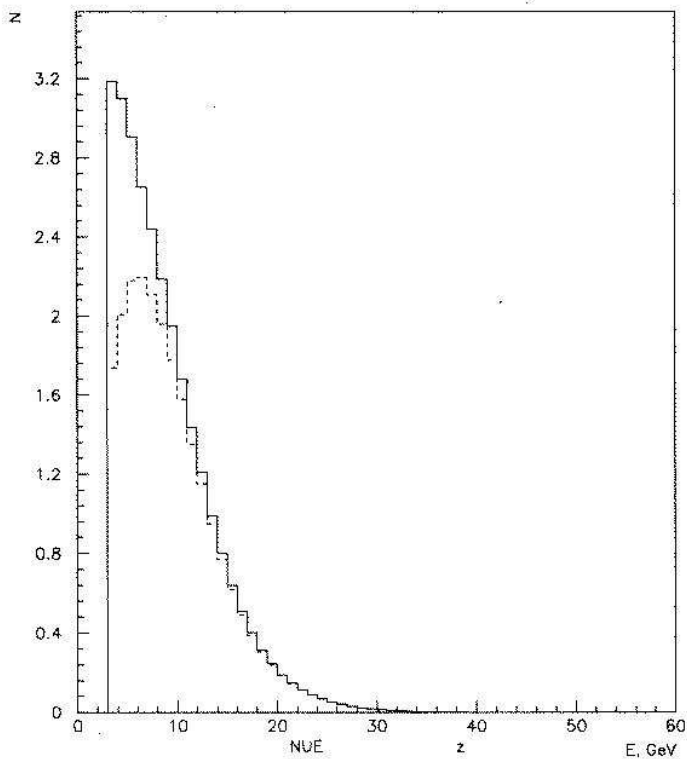,width=100mm}}
     \caption{$\nu_e$ spectrum in the experiment with focusing.
              Dashed line - the same without taking into account
              decays of muons.}
      \label{fig:foc}
   \end{center}
\end{figure}

\hspace*{0.5cm}
Here some difficulty was to introduce into the program the parabolic
lenses used in IHEP because there is no paraboloid shape in GEANT.
If define these lenses as cylinders, the magnetic field in them is
not only non-uniform but non-contiguous and GEANT does not trace
particles correctly. For this reason a routine for the step limitation
in the lenses was introduced. The step is limited so that it is either
entirely in a region with field or entirely in a region without field
or is very small. \\
\hspace*{0.5cm}
The electron neutrino spectrum with a contribution from decays of
muons is shown in Fig.~\ref{fig:foc}. \\
\hspace*{0.5cm}
The systematic uncertainty of the neutrino flux calculation consists
of the same uncertainties as for the beam with the focusing system off
and additional uncertaintied of the focusing system such as a possible
inaccurate target and lenses positioning and an uncertainty in the
simulation of starting points of particles inside the target. No
studies of the focusing system uncertainties were made with
this program, but the previous studies \cite{foc} show that, taking
into account muon flux measurements, they can be kept as low as 7\%

\section{The neutrino beam with a short decay base.}

\hspace*{0.5cm}
For this experiment \cite{short}, with rather complicated and asymmetrical
geometry of the decay region, it was very convenient to use GEANT tools for
the description of the geometry. \\
\hspace*{0.5cm}
The decay cavity can be described as 4 adjacent
(with steps in the vertical projection) trapezoids, asymmetric in the
vertical projection. It is narrow near the target to provide a
better radiation shielding and wider near the end closest to the detector.
It was made asymmetric because of the danger
of "underground" muons that can go around the iron muon filter, in the
earth under it due to a multiple scattering. \\
\hspace*{0.5cm}
In this experiment a contribution from muon decays is also rather
significant, including decays of muons in the muon filter (the path
length of muons in it is comparable or even bigger than the length of
the decay region).

\section{Conclusion}

\hspace*{0.5cm}
A program for the neutrino spectra calculations based on the
GEANT library is created. The range of possible applications of
the program is very wide.

\section{Acknowledgements}
\label{sec:acknowledgements}

\hspace*{0.5cm}
The author expresses the gratitude to A.S.Vovenko, A.I.Mukhin,
Yu.M.Sviridov, Yu.M.Sapunov for helpful discussions, V.I.Kravtsov,
Yu.M.Sapunov, K.E.Shestermanov for provided routines. \\
\hspace*{0.5cm}
The developments of the program for the neutrino spectra calculations
have been possible thanks to the computer power and other support
offered by the Physics institute of the University of Urbino (Italy).


\newpage
\begin{thebibliography}{99}
%
\bibitem{geant}GEANT manual, CERN Program library Long Writeup
W5013, Copyright CERN, Geneva, 1993
\bibitem{yields} Bojko N.I. et al., Yadernaya Fizika, v.31, p. 1246;
v.31, p. 1494 (1980); Preprint IFVE 79-77, Serpukhov, 1979 (Russian).
The numerical data are only in the preprint and in the IHEP data base
\bibitem{k0fnal} Scubic P. et al., Phys.Rev D18, p.3115 (1978)
\bibitem{kcern} Atherton H.W. et al., CERN 80-07 (1980)
\bibitem{method} Taylor F.E. et al., Phys.Rev. D14, p.1217 (1976)
\bibitem{utod} Johnson J.R. et al., Phys.Rev D17, p.1292 (1978)
\bibitem{bd} Blumlein J. et al., Phys. Lett. B279, p.405 (1991)
\bibitem{mufl} Bugorsky A.P. et al., NIM 146, p.367
\bibitem{foc} Bugorsky A.P. et al., preprint IFVE 78-116 (Russian),
Serpukhov, 1978; Belkov et al., preprint IFVE 82-99 (Russian),
Serpukhov, 1982
\bibitem{short} A.A. Borisov et al., Phys.Lett. B 369(1996) p.39.
%
\end{thebibliography}
\end{document}